\documentclass{emulateapj-rtx4}
\usepackage{apjfonts}

\input psfig.tex

\newcommand{\BI}{$B\!-\!I$}

\newcommand{\gz}{$g\!-\!z$}

\newcommand{\ZH}{[$Z$/H]}
\newcommand{\aFe}{[$\alpha$/Fe]}
\newcommand{\kms}{km\,s$^{-1}$} 
\newcommand{\Modot}{M_{\odot}}
\newcommand{\Msun}{$M_{\odot}$}
\newcommand{\Mcl}{{\cal{M}}_{\rm cl}}

\newcommand{\Rgal}{R_{\rm gal}}
\newcommand{\cM}{{\cal{M}}}
\newcommand{\ML}{${\cal{M}}/L$}
\newcommand{\mybibitem}[3]{\bibitem[{#1}({#2})]{#3}}
\newcommand{\mybibthree}[4]{\bibitem[{#2}({#3}){#1}]{#4}}
\newcommand{\picplace}[1]{\vbox{\hrule\@height 0.4pt\@width\hsize
\hbox to\hsize{\vrule\@width 0.4pt\@height#1\hfil
\vrule\@width 0.4pt\@height#1}\hrule\@height 0.4pt\@width\hsize}}

\slugcomment{Published as ApJ, {\bf 762}, 107 (2013)}

\shorttitle{New Evidence for a Bottom-Heavy IMF in Giant Elliptical Galaxies}
\shortauthors{P. Goudfrooij \& J. M. D. Kruijssen}

\begin{document}


\title{The Optical Colors of Giant Elliptical Galaxies and their
  Metal-Rich Globular Clusters Indicate a Bottom-Heavy Initial Mass Function}   


\author{Paul Goudfrooij\altaffilmark{1} and J. M. Diederik Kruijssen\altaffilmark{2}}


\altaffiltext{1}{Space Telescope Science Institute, 3700 San Martin
  Drive, Baltimore, MD 21218, USA; goudfroo@stsci.edu} 
\altaffiltext{2}{Max-Planck Institut f\"ur Astrophysik,
  Karl-Schwarzschild-Str.\ 1, D-85741 Garching, Germany;
  kruijssen@mpa-garching.mpg.de}





\begin{abstract}
We report a systematic and statistically significant offset between the
optical (\gz\ or \BI) colors of seven massive elliptical galaxies and the mean
colors of their associated massive metal-rich globular clusters (GCs) in the
sense that the parent galaxies are redder by $\sim$\,0.12\,--\,0.20 mag at a given
galactocentric distance. However, spectroscopic indices in the blue indicate
that the luminosity-weighted ages and metallicities of such galaxies are equal
to that of their averaged massive metal-rich GCs at a given galactocentric
distance, to within small uncertainties. 
The observed color differences between the red GC systems and their parent
galaxies cannot be explained by the presence of multiple stellar generations
in massive metal-rich GCs, as the impact of the latter to the populations'
integrated \gz\ or \BI\ colors is found to be negligible. However, we show that 
this paradox can be explained if the stellar initial mass function (IMF)
in these massive elliptical galaxies was significantly steeper at subsolar
masses than canonical IMFs derived from star counts in the solar neighborhood,
with the GC colors having become bluer due to dynamical evolution, causing 
a significant flattening of the stellar MF of the average surviving GC. 
\end{abstract}


\keywords{galaxies: elliptical and lenticular, cD --- galaxies: stellar
  content --- galaxies: formation --- galaxies: star clusters: general
  --- globular clusters: general}




\section{Introduction}              \label{s:intro}

The shape of the stellar initial mass function (IMF) is a 
fundamental property for studies of star and galaxy formation. It
constitutes a crucial assumption when deriving physical
parameters from observations. However, its origin remains relatively poorly
understood.  
A particularly important question is whether or not the IMF is ``universal''
among different environments, since galactic mass-to-light (\ML) ratios are
very sensitive to variations in the shape of the IMF at subsolar
masses. Several observational studies find that the IMF in various environments
within our Galaxy and the Magellanic Clouds seems remarkably uniform down to
detection limits of $\sim 0.1 - 1\, \Modot$ 
\citep[][and references therein]{krou01,chab03,bast+10}. 
However, recent studies indicate that the IMF may be different in more extreme  
environments: strong gravity-sensitive features in near-IR spectra of luminous
elliptical galaxies by \citet{vancon10,vancon11} favor a bottom-heavy
IMF ($\alpha \simeq -3$ in $dN/d\cM \propto \cM^{\alpha}$). If confirmed,
this result would have widespread implications on studies of stellar populations
and galaxy evolution. For example, such a bottom-heavy IMF would increase the
\ML\ ratio in the $K$ band by a factor $\sim$\,3 relative to a ``standard''
\citet{krou01} IMF, and hence invalidate the widespread assumption that \ML\
ratios in the near-IR are highly independent of galaxy type or mass. 

In this paper, we describe an effort to find independent evidence to confirm
or deny the presence of a bottom-heavy IMF in nearby giant elliptical
galaxies. We use observed optical colors and spectral line indices of
metal-rich globular clusters (GCs) and their parent galaxies in conjunction
with dynamical evolution modeling of GCs.  

Infrared studies of star formation within molecular clouds have shown that 
stars form in clusters or unbound associations with initial masses $\cM_{\rm
  cl,\,0}$ in the range $10^2 - 10^8$ \Msun\ 
\citep[e.g.,][and references therein]{ladlad03,pz+10,krui12}. 
While most star clusters with $\cM_{\rm cl,\,0} \la 10^4$ \Msun\ are thought to
disperse into the field population of galaxies within a few Gyr,
the surviving massive GCs constitute luminous compact sources that can
be observed out to distances of several tens of megaparsecs. Furthermore,
star clusters represent very good approximations of a ``simple stellar
population'' (hereafter SSP), i.e., a coeval population of stars with a
single metallicity, whereas the diffuse light of galaxies is typically
composed of a mixture of different populations.  
Thus, star clusters represent invaluable probes of the star formation rate
(SFR) and chemical enrichment occurring during the main star formation epochs
within a galaxy's assembly history.   

Deep imaging studies with the {\it Hubble Space Telescope (HST)\/} revealed
that giant elliptical galaxies typically contain rich GC systems with bimodal
color distributions 
\citep[e.g.,][]{kunwhi01,peng+06a}. Follow-up spectroscopy 
showed that both ``blue'' and ``red'' GC subpopulations are
nearly universally old, with ages $\ga$\,8 Gyr
\citep[e.g.,][]{cohe+03,puzi+05,brod+12}. 
This implies that the color bimodality is mainly due 
to differences in metallicity. 
The colors and spatial distributions of the blue GCs are usually 
consistent with those of metal-poor halo GCs in our Galaxy and M31,
while red GCs have colors and spatial distributions that are 
similar to those of the ``bulge'' light of their host galaxies
\citep[e.g.,][]{geis+96,rhozep01,bass+06,brostr06,peng+06a,goud+07}. 
Thus, the red metal-rich GCs are commonly considered to be physically
associated with the stellar body of giant elliptical galaxies. This
represents an assumption for the remainder of this paper.

\section{Galaxy Sample and GC Selection} \label{s:sample}

Our main galaxy selection criterion is the presence of a very clear bimodal
optical color distribution of its GCs as derived from high-quality, high
spatial-resolution (i.e., {\it HST}) imaging that allows GC detection
as close to the galaxy centers as possible. 
With this in mind, most galaxies in our sample were drawn from the ACS Virgo
Cluster Survey \citep[ACSVCS;][]{cote+04}, a homogeneous survey of 100
early-type galaxies and their GCs in the Virgo galaxy cluster, 
using the F475W and F850LP filters (hereafter $g$ and $z$,
respectively). 
Specifically, we use galaxy photometry from 
\citet{ferr+06} supplemented by GC data from
\citet{jord+09}. 
Galaxies were selected based on the properties of GC color
distributions as modeled by \citet{peng+06a} using the Kaye's Mixture Model
\citep[KMM;][]{mclbas88,ashm+94}, 
fitting two Gaussians to the data. 
Two criteria were used for selection: {\it (i)\/} the $p$-value of
the KMM fit needs to be $\leq$~0.01, and {\it (ii)\/} the color
distribution of the GCs needs to clearly show two distinct peaks as judged by
eye in Fig.~2 of \citet{peng+06a}. Finally, we exclude NGC 4365 
due to conflicting results on the presence
of intermediate-age GCs among its red GCs in the literature\footnote{See 
\citet{puzi+02}, \citet{lars+03}, and \citet{kund+05} versus
\citet{brod+05} and \citet{chie+11}.}. This selection yields the
high-luminosity galaxies NGC\,4472, NGC\,4486, NGC\,4649, NGC\,4552,
NGC\,4621, and NGC\,4473 (see Table~\ref{t:colgradfits}). 

In the context of the current study, it is important to include a comparison
between metallicity measurements of red GCs from colors versus from
spectral line indices (see Section~\ref{s:colgrads} below). As adequate line
index data for several red GCs is currently only available for one
galaxy in the ACSVCS sample, we add NGC 1407 to our galaxy sample 
given that high-quality {\it HST/ACS\/} imaging (in filters F435W (hereafter
$B$) and F814W ($I$)) as well as optical spectroscopy of both red GCs and
the diffuse light are available \citep{forb+06,cena+07,spol+08a,spol+08b}.  

The selection of red GCs in these galaxies was done by selecting GCs redder
than the color for which the KMM probability of a GC being member of the blue
versus the red peak is equal. In addition, we require GCs to be massive enough
to render the effect of stochastic fluctuations of the number of red giant
branch (RGB) and asymptotic giant branch (AGB) stars on the integrated colors
smaller than the typical photometric error of $\sigma(g\!-\!z)$ = 0.04
mag. Using the methodology of \citet{cerlur04} for an age of 12 Gyr and solar
metallicity\footnote{Fluctuation magnitudes for the $g$ and $z$ passbands are
  estimated by interpolating between those of $B$ and $V$ and $I_C$ and $J$,
  respectively, using the values in Table 5B of \citet{wort94}.}, this
requirement translates in a minimum cluster mass $\Mcl \simeq 2.5 \times
10^5 \:\Modot$. For galaxies in the Virgo cluster, this mass limit is
equivalent to selecting GCs with $z_{\rm AB} \leq 21.91$ mag under the
assumptions of {\it (i)\/} $\Mcl/L_z$ = 1.5 $\Modot/L_{z,\odot}$ for $Z =
Z_{\odot}$ \citep[this value varies only slightly with metallicity;][]{jord+07} 
and {\it (ii)\/} $m\!-\!M$ = 31.09 mag for the Virgo cluster 
\citep{jord+09}. The equivalent magnitude limit for the GCs in NGC 1407 is $I
\leq 23.28$ mag, using $\Mcl/L_I$ = 2.1 $\Modot/L_{I,\odot}$ for $Z = Z_{\odot}$
\citep{mara05} and $m\!-\!M$ = 31.60 mag \citep{forb+06}.

\section{Radial Color Gradients of Galaxies and their Red GC Systems}  \label{s:colgrads}

\begin{figure*}[ptbh]
\centerline{
\psfig{figure=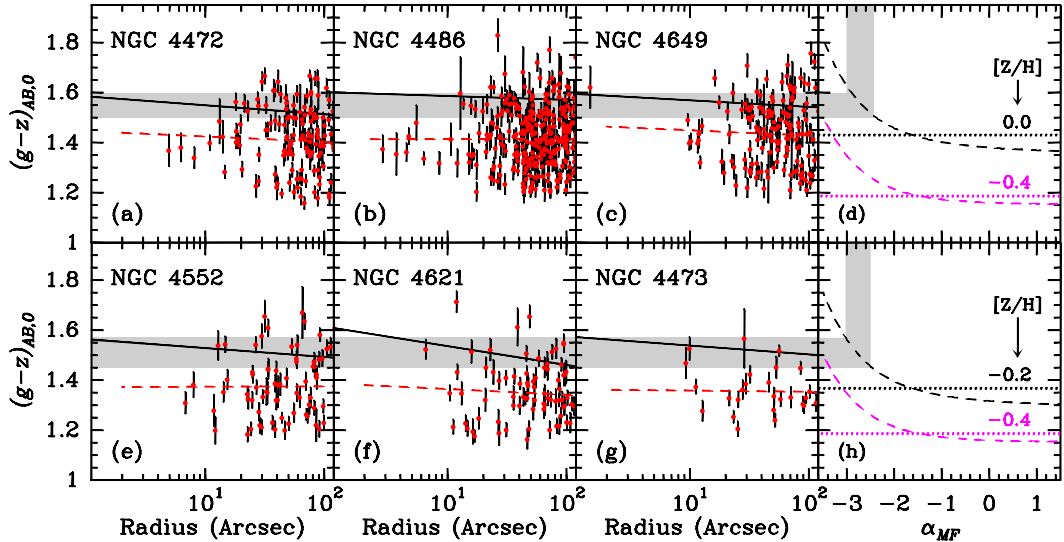,width=14.cm}
}
\caption{Radial color gradients of giant elliptical galaxies from the ACSVCS
  survey and their systems of massive red GCs. Panels (a)\,--\,(c) and (e)\,--\,(g) show
  dereddened $(g\!-\!z)_{\rm AB}$ versus projected galactocentric radius for
  the ACSVCS galaxies mentioned at the top left of the panel in
  question. Black solid lines represent linear least-square fits to the galaxy
  colors, red circles with black error bars represent individual red
  GCs, and the red dashed lines depict linear least-square fits to the
  red GC systems as a whole.  
  The dashed lines in panel (d) plot integrated $(g\!-\!z)_{\rm AB}$
  versus mass function (MF) slope $\alpha_{\it MF}$ for \citet{mari+08}
  isochrones for an age of 12 Gyr and the two \ZH\ values shown in the
  legend. The dotted lines in panel (d) indicate the colors for a
  \citet{krou01}  MF. Panel (h) is similar to panel (d), but now the upper
  (black) lines are drawn for \ZH\ = $-$0.2. The grey shaded region
  crossing panels (a) through (d) and (e) through (h) illustrate the
  range of MF slopes reproducing the galaxy colors for the upper
  isochrone lines drawn in panels (d) and (h), respectively. 
\label{f:rad_vs_col_ACSVCS}
}
\end{figure*}

\begin{figure}[ptbh]
\centerline{
\psfig{figure=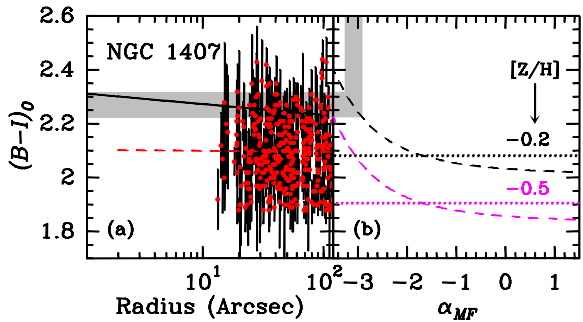,width=7.6cm}
}
\caption{Similar to Figure~\ref{f:rad_vs_col_ACSVCS}, but now for 
  NGC 1407 and its red GCs. Panel (a) shows $(B\!-\!I)_0$ versus projected
  galactocentric radius. 
  The dashed lines in panel (b) plot integrated $(B\!-\!I)_0$
  versus MF slope $\alpha_{\it MF}$ for \citet{mari+08} isochrones for an age
  of 12 Gyr and the two \ZH\ values shown in the legend. 
  Symbols, lines, and shading as in Figure~\ref{f:rad_vs_col_ACSVCS}.
\label{f:rad_vs_col_N1407}
}
\end{figure}

In this section we compare the colors of the red GCs (selected as
mentioned above) with those of the diffuse light of their parent
galaxies as a function of projected galactocentric radius (hereafter
$\Rgal$). All observed colors for the ACSVCS galaxies were dereddened using
the Galactic foreground reddening values listed in \citet{jord+09}. The
colors of GCs in NGC 1407 from \citet{forb+06} were already corrected for
Galactic foreground reddening. 
For each galaxy we calculate a mean color of the red GC population along with 
its radial gradient by means of a weighted linear least-squares fit between
the GC's colors {\it col\/} (i.e., \gz\ or \BI) and the logarithm of $\Rgal$ in arcsec:  
\begin{equation} 
 \overline{\mbox{\it col}} \; (\Rgal) = {\it col}_0 + 
 G_{\it col}  \: \log (\Rgal)
\label{eq:colgrad}
\end{equation}
where {\it col}$_0$ is the \gz\ or \BI\ color at log\,($\Rgal$) = 0 and
$G_{\it col}$ is the color gradient per dex in radius, $\Delta${\it
  (col)}/$\Delta \log(\Rgal)$. For each GC, uncertainties associated
with its measured color and its membership of the red population are taken
into account in the fit by adding the following two parameters in quadrature: 
{\it (i)\/} the photometric error of the GC's color, and {\it (ii)\/}
the inverse probability of the GC being a member of the red population ($p_{\rm
  red}$), as calculated by the KMM algorithm mentioned above, expressed in
magnitudes (i.e., log($1/p_{\rm red}$)\,/\,log(2.5)). 
Radial color profiles for the diffuse light of the parent galaxies were
determined using the same functional form as Eq.\ (\ref{eq:colgrad}). 
For NGC 1407, \BI\ data of the diffuse light were taken from
the surface photometry of \citet{spol+08a}, which was derived from the same
{\it HST/ACS\/} data as the GC photometry of \citet{forb+06}. ACSVCS galaxy
color gradients 
are taken from \citet{liu+11}, who used the surface photometry of
\citet{ferr+06}. Observed ACSVCS galaxy colors at $\Rgal = 1$ arcsec were
taken from \citet{ferr+06} and listed in Table~\ref{t:colgradfits}.   
Panels (a)\,--\,(c) and (e)\,--\,(g) of Figure~\ref{f:rad_vs_col_ACSVCS} show 
\gz\ versus $\Rgal$ for the 6 ACSVCS galaxies and their massive red GCs. 
Panel (a) of Figure~\ref{f:rad_vs_col_N1407} shows the equivalent for NGC 1407
and its red GCs. 
Note that while the radial gradients of the galaxies' colors are similar to
those of the ``average'' colors of their red GC systems, there is a
systematic offset between the two sets of radial color profiles in
the sense that the average color of red GCs is $\simeq$\,0.12\,--\,0.20 mag
bluer in \gz\ or \BI\ than that of their parent galaxies. This color
difference was already noted by \citet{peng+06a}, but they did not pursue
an investigation of its possible cause(s).  
The significance of this color difference is 7\,$\sigma$ to
22\,$\sigma$ depending on the galaxy, where $\sigma$ is the mean error
of the fit of eq.\ (\ref{eq:colgrad}) to the colors of red GCs in the
sample galaxies (see Table~\ref{t:colgradfits}).    

\begin{table}[tbh]
\footnotesize
\caption[]{Colors of sample galaxies and their massive red GCs.}
\label{t:colgradfits}
\begin{tabular*}{8.5cm}{@{\extracolsep{\fill}}cccccc@{}} \tableline \tableline
\multicolumn{3}{c}{~~} \\ [-2ex]  
NGC & $M_{V^0_T}$ & $(g\!-\!z)_{0,\,\rm gal}$ & $(g\!-\!z)_{0}$ & $G_{g\!-\!z}$ & rms \\
(1) & (2) & (3) & (4) & (5) & (6) \\ [0.5ex] \tableline
\multicolumn{3}{c}{~~} \\ [-1.5ex] 
4472 & $-$22.90 & 1.586 $\pm$ 0.005 & 1.425 & $-$0.021 & 0.010 \\
4486 & $-$22.66 & 1.598 $\pm$ 0.005 & 1.413 & $+$0.000 & 0.007 \\
4649 & $-$22.41 & 1.592 $\pm$ 0.005 & 1.449 & $-$0.022 & 0.010 \\ 
4552 & $-$21.36 & 1.558 $\pm$ 0.005 & 1.372 & $+$0.000 & 0.014 \\
4621 & $-$21.18 & 1.596 $\pm$ 0.005 & 1.365 & $-$0.023 & 0.013 \\
4473 & $-$20.70 & 1.563 $\pm$ 0.005 & 1.358 & $-$0.005 & 0.017 \\ [0.5ex] \tableline 
\multicolumn{3}{c}{~~} \\ [-1.8ex] 
 NGC & $M_{V^0_T}$ & $(B\!-\!I)_{0,\,\rm gal}$ & $(B\!-\!I)_{0}$ &
  $G_{B\!-\!I}$ & rms \\  [1.2ex] \tableline
\multicolumn{3}{c}{~~} \\ [-1.8ex] 
1407 & $-$21.86 & 2.313 $\pm$ 0.005 & 2.103 & $-$0.004 & 0.007 \\ [0.5ex] \tableline 
\multicolumn{5}{c}{~~} \\ [-1.2ex]
\end{tabular*}
\tablecomments{Column (1): NGC number of galaxy. Column (2): absolute
  $V$ magnitude of galaxy. Column (3:) ${\it col}_0$ of galaxy.
  Column (4): ${\it col}_0$ of red GCs. Column (5): color gradient
  $G_{\it col}$ of red GCs in mag/dex.  Column (6): mean rms error of fit of Eq.\
  (1) to colors of red GCs (in mag).} 
\end{table}

Under the assumption that surviving GCs represent the high-mass end of the
star formation process that also populated the field \citep{elmefr97}, this color
difference is difficult to understand in terms of a metallicity
difference. While we deem it likely that the spread of colors exhibited by the
red GC systems in Figure~\ref{f:rad_vs_col_ACSVCS} is 
due (at least in part) to a range of GC metallicities (and/or ages), one would
expect that the same range of metallicities (and/or ages) be present among the
field stars and hence that the {\it average\/} colors would be similar. 

\begin{figure}[tbph]
\centerline{
\psfig{figure=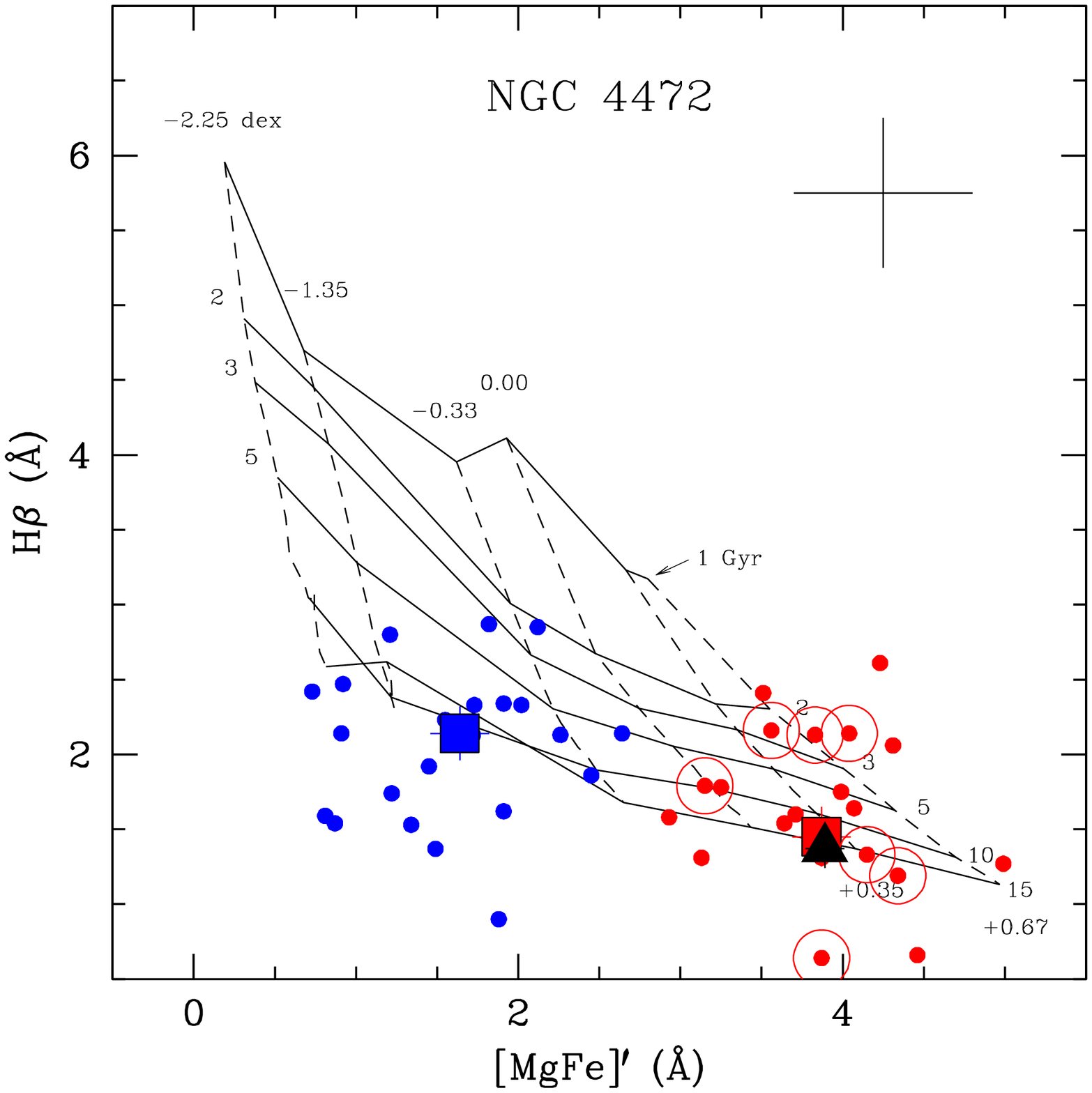,width=7.5cm}
}
\caption{H$\beta$ versus [MgFe]$'$ for GCs in NGC~4472 from the data of 
  \citet{cohe+03}. Small blue and red circles indicate GCs designated by
  \citet{geis+96} as ``blue'' and ``red'', respectively. Large filled blue and
  red squares represent weighted average values of H$\beta$ and [MgFe]$'$ of
  the blue and red GCs, respectively. Large open red circles indicate red GCs
  with HST/ACS photometry (see Figure~\ref{f:spec_vs_phot}a and
  Sect.~\ref{s:mismatch}). Error bars at the top right
  indicate typical uncertainties for individual GCs. SSP models of
  \citet{thom+03} are overplotted for [$\alpha$/Fe] = +0.3. Dashed lines
  indicate \ZH\ values of $-$2.25, $-$1.35, $-$0.33, 0.00, 0.35, and 0.67 dex. 
  Solid lines indicate ages of 1, 2, 3, 5, 10, and 15 Gyr.
  For comparison, the black triangle represents H$\beta$
  and [MgFe]$'$ for the diffuse light of NGC~4472 at $\Rgal = 50''$. See
  Sect.\ \ref{s:colgrads}. 
\label{f:M49specplot}
}
\end{figure}

\begin{figure}[tbph]
\centerline{
\psfig{figure=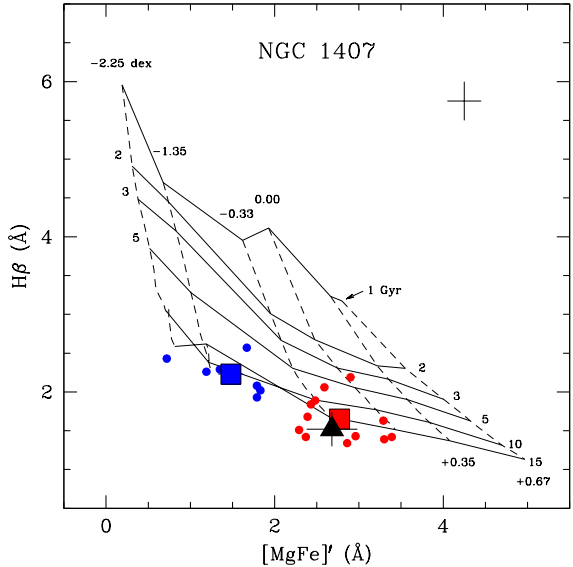,width=7.5cm}
}
\caption{Similar to Figure~\ref{f:M49specplot}, but now for GCs in NGC~1407
  from the data of \citet{cena+07} and for the diffuse light of NGC~1407 at
  $\Rgal = 50''$ from the data of \citet{spol+08b}. Large open red circles
  are not shown in this case since all GCs shown have HST/ACS photometry (cf.\ 
  Figure~\ref{f:spec_vs_phot}b).  
\label{f:N1407specplot}
}
\end{figure}

This argument is corroborated by spectroscopic Lick index data for red GCs
and the diffuse light in the two giant elliptical galaxies in our
sample (or in the literature at large) for which we found such data
available in the literature for both a significant number of red GCs
and the diffuse light within a range in $\Rgal$ sampled by both
components. 
As to the GC spectra, \citet{cohe+03} obtained Keck/LRIS spectra of 47
GCs in NGC~4472, while \citet{cena+07} used Keck/LRIS to obtain
spectra for 20 GCs in NGC~1407. In both studies, the resulting GCs 
were split approximately evenly between blue and red GCs. To evaluate
spectroscopic ages and metallicities, we use the indices H$\beta$ and
[MgFe]$' = \left( \mbox{Mg}\,b \times (0.72\, \mbox{Fe}_{5270} +
  0.28\, \mbox{Fe}_{5335}) \right)^{1/2}$, 
respectively, and compare them to predictions of the SSP models of
\citet{thom+03}. 
The [MgFe]$'$ metallicity index was chosen because \citet{thom+03}
found it to be independent of variations in [$\alpha$/Fe] ratio. This is
relevant since giant elliptical galaxies and their metal-rich GCs are known to
exhibit supersolar $\alpha$/Fe ratios \citep[e.g.,][]{trag+00,puzi+06}. 
Figure~\ref{f:M49specplot} shows 
H$\beta$ versus [MgFe]$'$ for the GCs in NGC 4472 from \citet{cohe+03},
subdivided into blue and red GCs. 
As the uncertainties are significant for the individual GCs, we also
indicate weighted average indices for the blue and red GCs (see large blue and 
red symbols in Figure~\ref{f:M49specplot}). 
Note that \ZH\ $\simeq +0.30$ for the average red GC in
Figure~\ref{f:M49specplot}, which corresponds to \ZH\ = $-0.07$ on the
\citet{zinwes84} scale \citep{cohe+03}.   
For comparison, we overplot H$\beta$ and [MgFe]$'$ for the diffuse light of
NGC~4472 \citep{davi+93,fish+95} at $\Rgal$ = 50$''$. Note that the median
$\Rgal$ of the red GCs targeted by \citet{cohe+03} is 59\farcs1.  
Similarly, Figure~\ref{f:N1407specplot} shows H$\beta$ versus [MgFe]$'$ for 
the blue and red GCs in NGC 1407 from \citet{cena+07} and the diffuse light of
NGC 1407 at $\Rgal$ = 50$''$ from \citet{spol+08b}. For comparison, the median
$\Rgal$ of the red GCs targeted by \citet{cena+07} is 50\farcs3. 
We conclude that the average
red GC and the diffuse light have equal values of H$\beta$ and 
[MgFe]$'$ to within 1$\sigma$ for both NGC 1407 and NGC 4472, the two
galaxies for which we could find spectroscopic ages and metallicities for both red
GCs and the diffuse light, at comparable galactocentric distances, in the
literature. This indicates that the mean ages and metallicities of the red GCs
and the underlying diffuse light of giant elliptical galaxies are indeed the
same within the uncertainties.    

We thus arrive at a situation where the mean spectroscopic ages and
metallicities of the diffuse light of giant elliptical galaxies and their red
GCs seem consistent with one another while the \gz\ or \BI\ colors of the
diffuse light of the galaxies are systematically redder than those of their
average red GC. We evaluate three possible solutions to this paradox below.  

\section{Possible Causes of the Color Offset Between Red GC Systems
  and  Their Parent Galaxies} \label{s:causes}

\subsection{A Mismatch between Photometric and Spectroscopic
  Samples?} \label{s:mismatch}

\begin{figure}[tbh]
\centerline{
\psfig{figure=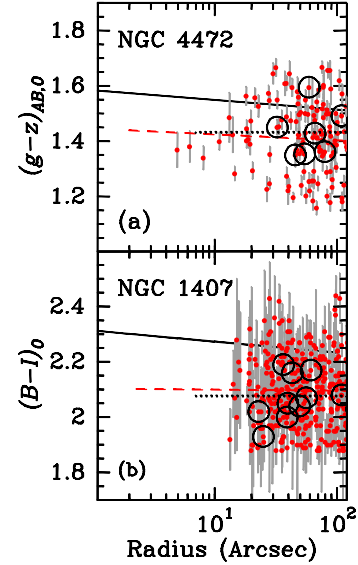,width=4.5cm}
}
\caption{Panels (a) and (b) are near-copies of
  Figures~\ref{f:rad_vs_col_ACSVCS}a and \ref{f:rad_vs_col_N1407}a,
  respectively, except that error bars are now shown in grey. Large open 
  circles in panels (a) and (b) identify the red GCs in the HST data for
  which Lick index data were shown in Figures~\ref{f:M49specplot} and
  \ref{f:N1407specplot}, respectively. The black dotted lines indicate the
  mean colors of the two sets of red GCs shown with large open circles. Note
  that the black dotted lines almost coincide with the red dashed lines, which
  depict linear fits to Eq.~\ref{eq:colgrad} for the {\it full\/} red GC
  systems.  
\label{f:spec_vs_phot}
}
\end{figure}

As a first possible solution to the paradox described in the previous Section, we
consider the hypothesis that the red GCs in NGC~4472 and NGC~1407 
whose spectroscopic Lick index data were shown in
Figures~\ref{f:M49specplot} and \ref{f:N1407specplot}, respectively, happen to be
redder and more metal-rich on average than the respective full photometric
samples of red GCs. This hypothesis is rejected by the data, as illustrated
in Figure~\ref{f:spec_vs_phot} in which open circles highlight the red GCs
that have counterparts in Figures~\ref{f:M49specplot} and
\ref{f:N1407specplot}. Dotted black lines indicate the mean \gz\ or \BI\
colors of the GCs with Lick index data, which are consistent with the mean
colors of the {\it full\/} red GC systems to within 1\,$\sigma$. Recalling 
that the significance of the observed color differences between the metal-rich
GC systems and their host galaxies is in the range 7\,--\,22 $\sigma$, we
conclude that the red GCs with Lick index data are not misrepresenting the 
full systems of red GCs in terms of their \gz\ or \BI\ colors. 

\subsection{An Effect of the Presence of Multiple Stellar Populations
  in Massive GCs?} \label{s:multpops}

Several detailed photometric and spectroscopic investigations over the last
decade have established that massive Galactic GCs typically host multiple,
approximately coeval, stellar populations \citep[see the recent review by][and
references therein]{grat+12}. The recognition of this fact came
mainly with the common presence of anticorrelations between
[Na/Fe] and [O/Fe] among individual stars \citep[often dubbed ``Na-O
anticorrelations''; see, e.g.,][]{carr+10}, which are seen within virtually
all Galactic GCs studied using multi-object spectroscopy with 8-m-class
telescopes to date. The leading theory on the cause of these abundance
variations is the presence of a second generation of stars formed out of
enriched material lost (with low outflow velocity) by ``polluters'' (massive
stars and/or intermediate-mass AGB stars) of the first
generation. High-temperature proton-capture nucleosynthesis in the
atmospheres of the ``polluters'' increased the abundances of He, N, and Na
relative to Fe \citep[e.g.,][]{renz08}.  

Note that this theory only predicts observable star-to-star abundance
variations among {\it light\/} elements (up to $^{13}$Al), which is
consistent with what is seen among almost all Galactic GCs. 
Variations in {\it heavy\/} (e.g., Fe-peak) element abundances have only been
found within two very massive {\it metal-poor\/} GCs: $\omega$\,Cen
\citep[e.g.][]{lee+05,johpil10} and M\,22 \citep{marino+09}. It is
thought that such GCs represent surviving nuclei of accreted dwarf galaxies
whose escape velocities were high enough to retain Fe-peak elements from SN
explosions \citep[e.g.,][]{lee+07,geor+09b}. However, we deem this accretion
scenario rather unlikely for the formation history of {\it red\/} GCs given
their high metallicities. We therefore restrict the following discussion to
GCs without a spread in iron abundance. 

The effect of light-element abundance variations within GCs to {\it photometry\/}
seems to be more subtle than that to spectroscopic line indices. While
photometric splitting of RGB, main sequence (MS), and/or subgiant
branch (SGB) sequences have been observed in color-magnitude diagrams
(CMDs) of several Galactic GCs, such splits only show up in CMDs that
involve filters shortward of $\sim$\,4000 \AA, while they disappear
when using visual passbands such as $B$, $V$, and/or $I$ \citep[see,
e.g.,][]{marino+08,krav+11,milo+12}.   

To evaluate the effect of the presence of multiple stellar populations in
massive GCs to their colors, we determine colors of various stellar types
from synthetic spectra with chemical compositions typical of first-- and
second-generation stars found in massive GCs, and we use the resulting
stellar spectra to calculate integrated colors of the full stellar
population. This procedure is described in detail in Appendix
\ref{s:appendixA}. As shown there, the expected impact of the presence of two
stellar generations to the integrated \BI\ and \gz\ colors of metal-rich GCs
is offsets of $-$0.021 and $-$0.027 mag, respectively, relative to a true
SSP. As this is only a small fraction of the observed color offsets between
red GC systems and the diffuse light of giant elliptical galaxies, it
seems that an additional effect is needed to produce that color offset.

\subsection{A Steep Stellar Mass Function
  Slope in the Field Star Component?} \label{s:color_vs_alpha} 

At a given age and chemical composition, the only fundamental property of
a simple stellar population that can affect its integrated color (or spectral
line index) significantly is the shape of its stellar MF. 
Figures~\ref{f:rad_vs_col_ACSVCS}d, \ref{f:rad_vs_col_ACSVCS}h, and
\ref{f:rad_vs_col_N1407}b illustrate the effect of changing 
the slope $\alpha$ of the stellar MF to the integrated \gz\ or \BI\ color of a
stellar population. The dashed curves in these panels were determined from Padova
isochrones \citep{mari+08} of age 12 Gyr and for the \ZH\ values shown in
the figure. After rebinning the isochrone tables to a uniform bin size in the
stellar mass $\cM_{\star}$ using linear interpolation, weighted 
luminosities of the population were derived for the relevant 
passbands by weighting individual stellar luminosities by a factor
$(\cM_{\star}/\cM_{\rm max})^{\alpha}$ where  
$\cM_{\rm max}$ is the maximum stellar mass reached in the isochrone
table. For reference, the dotted lines show the expected colors for a
``standard'' \citet{krou01} IMF.  

\begin{figure}[tbh]
\centerline{
\psfig{figure=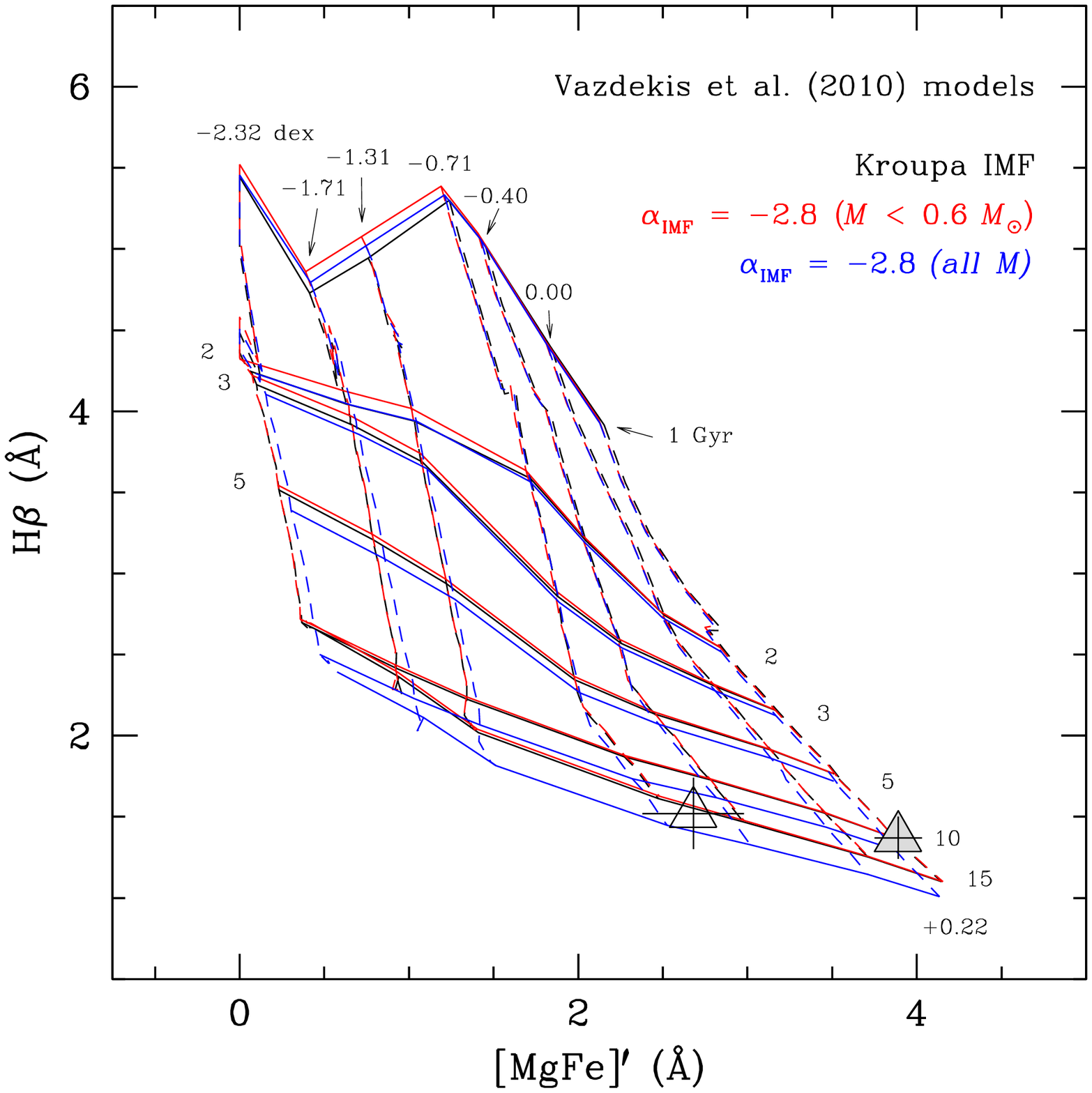,width=8cm}
}
\caption{H$\beta$ versus [MgFe]$'$ for SSP model predictions of
  \citet{vazd+10} using different IMFs (see legend on the right). Black lines: 
  \citet{krou01} IMF. Red lines: $\alpha = -2.8$ for $\cM \leq 0.6\; M_{\odot}$ and
   Salpeter IMF for $\cM > 0.6 \; M_{\odot}$. Blue lines: $\alpha = -2.8$ for all
   stellar masses. Dashed lines 
  indicate \ZH\ values of $-$2.32, $-$1.71, $-$1.31, $-$0.71, $-$0.40, 0.00,
  and 0.22 dex. Solid lines indicate ages of 1, 2, 3, 5, 10, and 15 Gyr.
  For comparison, the grey and open triangles represent H$\beta$
  and [MgFe]$'$ at $\Rgal = 50''$ for the diffuse light of NGC~4472 and
  NGC~1407, respectively. See the discussion in Sect.\ \ref{s:color_vs_alpha}. 
\label{f:V10_IMFplot}
}
\end{figure}

A comparison of panels (a)\,--\,(c) with panel (d) of
Figure~\ref{f:rad_vs_col_ACSVCS} reveals several items of interest. First, the
``average'' colors of the red GC systems are consistent with the prediction of
a SSP model with age = 12 Gyr, \ZH\ = 0.0, and a Kroupa MF. 
This is also consistent with the spectroscopic age and \ZH\ found for the
average red GC found above in Section~\ref{s:colgrads}. 
Second, the redder colors of the parent galaxies with respect to the average
colors of the red GC systems can be explained if the MF of the field star
component is ``bottom-heavy'' ($-3.0 \la \alpha \la -2.4$ for \ZH\ = 0.0)
relative to that of the red GC system, especially in the inner regions. 
This assumes that the field stars and the massive red GCs share the same
distributions of age and metallicity, for which we showed spectroscopic
evidence in Figs.~\ref{f:M49specplot} and \ref{f:N1407specplot}. 
  In this regard it is important to note that the dependence of the
  blue-visual spectral indices H$\beta$ and [MgFe]$'$ to changes in the MF
  slope in the range indicated by the color differences between red GCs and
  their parent galaxies is negligible. This is illustrated in
  Figure~\ref{f:V10_IMFplot} where we compare the values of H$\beta$ and
  [MgFe]$'$ for NGC 4472 and NGC 1407 at $\Rgal$ = 50$''$ (cf.\ Figs.\
  \ref{f:M49specplot} and \ref{f:N1407specplot}) with predictions of the 
  \citet{vazd+10} SSP models which were calculated for a variety of IMF
  types and slopes\footnote{Note that \citet{vazd+10} defined IMF slopes 
    $\mu$ as $dN/d\cM \propto \cM^{-(\mu + 1)}$. Hence, our $\alpha = -(\mu
    +1)$.}. We plot model predictions for {\it (i)\/} a Kroupa IMF, {\it
    (ii)\/} an IMF with a Salpeter slope ($\alpha = -2.3$) for stellar masses
  $\cM > 0.6 \; M_{\odot}$ and $\alpha = -2.8$ for $\cM \leq 0.6 \;
  M_{\odot}$, and {\it (iii)\/} an IMF with $\alpha = -2.8$ for all stellar masses. 
  As Figure~\ref{f:V10_IMFplot} shows, the three sets of model predictions are
  consistent with one another to within the measurement uncertainties in the
  relevant region of parameter space. 

Similar conclusions can be drawn when comparing panels (e)\,--\,(g) with panel
(h) of Figure~\ref{f:rad_vs_col_ACSVCS} and panel (a) with panel (b)
of Figure~\ref{f:rad_vs_col_N1407}, except that the average colors of the
red GC systems of the less luminous galaxies depicted there (see
Table~\ref{t:colgradfits}) are better matched by a slightly lower
metallicity ($[Z/H] \simeq -0.2$). This is consistent with the
color-magnitude relation among elliptical galaxies when interpreting
color changes in terms of metallicity changes:\  
the slope in the $V\!-\!K$ vs.\ $V$ relation among E galaxies in the
Virgo galaxy cluster is $-0.079 \pm 0.007$ 
\citep{bowe+92}, 
which translates to a \ZH\ vs.\ galaxy magnitude relation of
$-$0.093 $\pm$ 0.008 dex/mag according to the \citet{mari+08} SSP models at an 
age of 12 Gyr.

A possible mechanism for producing the steep MFs in giant elliptical galaxies
(relative to those of their massive red GCs) indicated by our results is
discussed in the next Section.   

\section{Dynamical Evolution Effects on Stellar
  Mass Functions of Globular Clusters} \label{s:evol}

The gradual disruption of star clusters over time affects the shape of their 
stellar MF, because the escape probability of stars from their parent cluster
increases with decreasing stellar mass
\citep[e.g.,][]{baumak03}. 
This effect has been used to explain the observed stellar MFs in
ancient Galactic GCs, some of which are  flatter than canonical IMFs
\citep[e.g.,][]{dema+07,paus+10}.  
Such flat MFs in GCs can explain their low observed
mass-to-light ratios when compared with predictions of SSP
models that use canonical IMFs \citep[e.g.,][]{krumie09}. 
The flattening of the MFs in Galactic GCs is greatest for GCs
that are closest to dissolution \citep{baum+08,krui09}. 

Here, we explore the hypothesis that the IMF in the (high-metallicity, 
and likely high-density) environments that prevailed during the main
star formation events that formed the stars in the bulge and the red GCs
of giant elliptical galaxies was relatively steep ($-3.0 \la \alpha
\la -2.4$, as indicated by the analysis in the previous Section),
i.e., steeper than the canonical IMF seen in the solar neighborhood, 
and that dynamical evolution of the star clusters 
formed during those events caused the stellar MF of the ``average''
surviving massive star cluster to be similar to a canonical IMF. 
To this end, we use the GC evolution model of \citet[][hereafter
K09]{krui09} which incorporates the effects of stellar evolution, stellar
remnant retention, dynamical evolution in a tidal field, and mass
segregation. For the purposes of the current study, we select K09 models
that {\it (i)\/} feature solar metallicity and ``default'' kick velocities of
stellar remnants (white dwarfs, neutron stars, and black holes), and {\it
  (ii)\/} produce GCs with masses $\cM_{\rm cl} \geq 2.5 \times 10^5 \:
\Modot$ at an age of 12 Gyr, i.e., the GCs shown in
Figures~\ref{f:rad_vs_col_ACSVCS} and \ref{f:rad_vs_col_N1407} 
 (cf.\ Section \ref{s:sample}). We further consider \citet{king66}
profiles with $W_0$ values of 5 and 7 and environmentally dependent cluster 
dissolution time scales $t_0$\footnote{$t_0$ is defined by $t_{\rm dis} = t_0
  \, \cM_{\rm cl}^{\gamma}$ where $t_{\rm dis}$ is the cluster disruption time and
$\gamma$ sets the mass dependence of cluster disruption. 
See K09 for details.} of 0.3, 0.6, 1.0, and 3.0
Myr. For reference, $t_0 = 1.3$ Myr yields a reasonable fit to the globular
cluster mass function of all (surviving) Galactic GCs \citep{krupor09}. 

Note however that GC systems in any given
galaxy are expected to exhibit a range of $t_0$ values. For example,
disruption rates of GCs in dense high-redshift environments were likely higher
than in current quiescent galaxies \citep{krui+12} due to the enhanced
strength and rate of tidal perturbations. Likewise, GCs on eccentric orbits
with smaller perigalactic distances $R_{\rm peri}$ should have smaller $t_0$
values than otherwise (initially) similar GCs with larger values of
$R_{\rm peri}$. 
Finally, the rate of mass loss from GCs by evaporation,  
$\mu_{\rm ev}$, effectively scales with the mean GC half-mass density
$\rho_{\rm   h}$ as $\mu_{\rm ev} \propto \rho_{\rm h}^{0.5}$ for tidally
limited GCs (e.g., \citealt{mclfal08,giel+11,goud12}). Among GCs for which 
two-body relaxation has been the dominant mass loss mechanism, 
high-$\rho_{\rm h}$ GCs thus feature smaller $t_0$ values than
low-$\rho_{\rm h}$ GCs at a given $\cM_{\rm cl}$. 

\begin{figure}[tbhp]
\centerline{
\psfig{figure=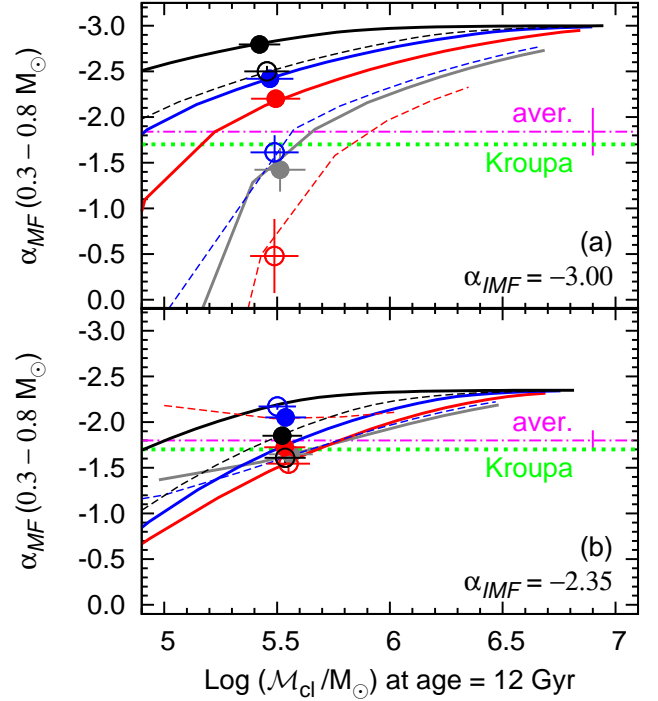,width=8.3cm,angle=-90.}
}
\caption{{\it Panel (a)}: MF slope $\alpha_{\it MF}$ in the range $0.3 <
  \cM_{\star}/\Modot < 0.8$ versus GC mass at an age of 12 Gyr for
  $\alpha_{\it IMF} = -3.00$ according to solar metallicity models of
  K09 that produce surviving GCs with $\cM_{\rm cl} \geq 2.5 \times
  10^5 \: \Modot$ at an age of 12 Gyr. Solid lines represent models with
  \citet{king66} parameter $W_0 = 7$, while dashed lines do so for $W_0 =
  5$. From top to bottom, the lines represent models with $t_0$ = 3.0, 1.0,
  0.6, and 0.3 Myr (the latter only for $W_0 = 7$). Circles indicate
  weighted mean values for the abscissa and ordinate for each model, assuming
  a \citet{sche76} initial cluster mass function to assign weights and 
  cluster masses $2.5 \times 10^5 \leq \cM_{\rm cl}/\Modot \leq 10^7$
  at an age of 12 Gyr.  
  The magenta dash-dotted line (along with an 
  error bar) indicates the weighted average $\alpha_{\it MF}$ of all 
  circles (using inverse variances as weights), while the green dotted line
  indicates the Kroupa IMF.  
  {\it Panel (b)}: Similar to panel (a), but now for a \citet{salp55} IMF
  ($\alpha_{\it IMF} = -2.35$). 
\label{f:alpha_vs_Mcurr}
}
\end{figure}

Figure~\ref{f:alpha_vs_Mcurr} shows the present-day MF slopes $\alpha_{\it
  MF}$ in the stellar 
mass range 0.3\,--\,0.8 \Msun\ 
according to the K09 models for surviving GCs as a function of $\cM_{\rm
  cl}$. The two panels (a) and (b) show the results for two different slopes
of the stellar IMF: $\alpha_{\it IMF} = -3.00$ and $-2.35$, respectively. We
also indicate ``weighted mean'' values of $\alpha_{\it MF}$ for a GC
population with $\cM_{\rm cl} \geq 2.5 \times 10^5 \; \Modot$ at an age of
12 Gyr that was formed with a \citet{sche76} initial cluster mass function and
a maximum initial mass of $10^7\:\Modot$. For reference, a power-law fit
to the Kroupa IMF in the mass range 0.3\,--\,0.8 \Msun\ yields
$\alpha_{\it MF} = -1.7$ (cf.\ panels (d) and (h) of
Fig.~\ref{f:rad_vs_col_ACSVCS}).   

Focusing on the ``weighted mean'' values of $\alpha_{\it MF}$ for the
K09 models shown in Figure~\ref{f:alpha_vs_Mcurr}, we find overall
mean values $\overline{\alpha_{\it MF}} = -1.84 \pm 0.26$ for
$\alpha_{\it IMF} = -3.00$ and $\overline{\alpha_{\it MF}} = -1.80 \pm
0.11$ for $\alpha_{\it IMF} = -2.35$, indicating that GCs with a steeper IMF
experience a stronger evolution of the stellar MF. Note that {\it both of\/}
these values are consistent with a Kroupa MF and hence consistent with the
observed mean colors of red GCs shown in Figure~\ref{f:rad_vs_col_ACSVCS}. 
It is important to realize that the stronger flattening of $\alpha_{\it MF}$
over time in massive GCs with steeper IMFs
is mainly caused by {\it the weaker effect of
  retained massive stellar remnants on the escape rate of massive stars in GCs
  with steeper IMFs relative to that in GCs with flatter IMFs}.  
As described in detail in K09, retained massive
stellar remnants boost the escape rate of massive stars relative
to that of low-mass stars. 
For a Kroupa IMF, this effect limits the 
flattening rate of $\alpha_{\it MF}$ over time relative to GCs with a lower
fraction of massive stellar remnants (see K09). However, GCs with steeper IMFs
feature a smaller mass fraction taken up by massive remnants. This notably
reduces the `dampening effect' of retained massive remnants on the evolution of
$\alpha_{\it MF}$.  
As a result, the present-day stellar MFs in GCs are consistent with any IMF
slope $\alpha \la -1.7$. This degeneracy 
implies that the IMF cannot be uniquely constrained in this range for
individual GCs. 
However, the large {\it spread\/} of observed MF slopes among 
Galactic GCs \citep{dema+07,paus+10} seems to be more consistent with
a bottom-heavy IMF (compare both panels in Figure~\ref{f:alpha_vs_Mcurr}). 

We conclude that the difference in \gz\ or \BI\ color between giant
elliptical galaxies and their ``average'' massive red GCs shown in 
Figures~\ref{f:rad_vs_col_ACSVCS} and \ref{f:rad_vs_col_N1407} is consistent
with long-term dynamical evolution of star clusters that were formed with a
steeper IMF than those derived from star counts in the solar neighborhood. 
Among the three mechanisms we think could conceivably
cause this color difference, this is the only one that seems
consistent with the data. Hence we suggest that this color difference constitutes
evidence for a ``bottom-heavy'' IMF in luminous elliptical galaxies.

\section{Discussion}  \label{s:disc}

Our results add to a developing consensus that luminous elliptical galaxies
have ``heavy'' IMFs, with a greater contribution from low-mass dwarf stars
than IMFs determined from star count studies in the Milky Way. 
\citet{cena+03} studied the Ca\,{\sc ii} triplet near 8600 \AA, which is
stronger in RGB stars than in MS stars, in a sample of elliptical
galaxies. They found that the strength of the Ca\,{\sc ii} triplet decreases
with increasing velocity dispersion among elliptical galaxies. This is hard to
explain in terms of an Ca abundance trend since Ca is an $\alpha$
element, and [$\alpha$/Fe] as derived from Mg and Fe features near 5200 \AA\
does not show this trend. Hence, they suggested that this trend might
reflect an increasing IMF slope with increasing galaxy mass, as indicated by
SSP models of \citet{vazd+03}. Similarly, \citet{vancon10,vancon11} found
strong equivalent widths of the Na\,{\sc i}\,$\lambda\lambda 8183, 8195$
doublet and the molecular FeH band at 9916\,\AA\ in near-IR spectra of giant
elliptical galaxies. As both of those indices are stronger in red dwarf stars than
in red giants, their strongly favored explanation is the presence of a
significantly larger number of late M dwarf stars than that implied by a
Kroupa IMF, even though there is a slight degeneracy with [Na/Fe] ratio
\citep[see also][]{convan12b,smit+12}.  
Finally, \citet{capp+12} studied stellar \ML\ ratios [$(\cM/L)_{\rm
  stars}$] derived from integral-field spectroscopy of 260 
early-type galaxies. They found that the ratio of $(\cM/L)_{\rm stars}$ by the
\ML\ predicted by SSP models that use a Salpeter IMF $[(\cM/L)_{\rm Salp}]$
varies systematically with galaxy velocity dispersion: galaxies with 
the largest values of $(\cM/L)_{\rm stars}/(\cM/L)_{\rm Salp}$ have the highest
velocity dispersions \citep[see also][]{tort+12}. 
However, dynamical modeling such as that done by 
\citet{capp+12} cannot unambiguously constrain whether the highest values of
$(\cM/L)_{\rm stars}/(\cM/L)_{\rm Salp}$ are due to a large population of
low-mass stars (i.e., a bottom-heavy IMF) or large number of massive stellar
remnants (i.e., a top-heavy IMF). The same degeneracy is present for mass
estimates from gravitational lensing \citep[e.g.,][]{treu+10}. In contrast, our
results indicate that the bottom-heavy IMF is the more likely solution. 

Finally, we review the dependence of our results on SSP model ingredients. At
old ages ($\ga 10$ Gyr) and high metallicities (\ZH\ $\ga -0.5$), SSP models
still suffer from several limitations. The likely most significant 
{\it internal\/} uncertainty is related to lifetimes of AGB stars
\citep[see, e.g.,][]{melb+12}. In terms of external uncertainties, we note
that integrated colors and spectral indices of several SSP models are calibrated
to observations of the two Galactic bulge GCs NGC~6528 and NGC~6553 that have
$Z \simeq Z_{\odot}$ \citep[e.g.,][]{thom+03,mara05}. However, 
there are no published MFs for those two GCs available to our
knowledge, yielding an uncertainty in the absolute
calibration\footnote{If the present-day MF of those GCs is flatter than the
  Kroupa or Chabrier IMFs, as might be expected given the strong tidal field at their
  location deep in the Galaxy's potential well, their optical colors would be
  bluer than if they had Kroupa or Chabrier MF's (cf.\ Figures
  \ref{f:rad_vs_col_ACSVCS} and \ref{f:rad_vs_col_N1407}). Thus, absolute ages
  and/or [$Z$/H] values of old metal-rich populations derived from integrated
  colors would currently be slightly overestimated.}.   
Notwithstanding such limitations, our main result is
based on \gz\ or \BI\ color {\it differences\/} between the average red 
massive GC and their parent galaxies and the absence of such
differences in H$\beta$ and [MgFe]$'$ indices. As the interpretation
of {\it relative\/} colors or line strengths should be largely
model-independent, our result strongly suggests a bottom-heavy IMF in massive 
elliptical galaxies.

\acknowledgments
The authors acknowledge useful discussions with Tom Brown, Mike Fall, Harry
Ferguson, Jessica Lee, Guido de Marchi, and Thomas Puzia.  
We thank Fiorella Castelli for her help in compiling some routines within the
SYNTHE procedure.  
We appreciated the referee's swift review, including insightful
  comments which improved the presentation of some results. 
The SAO/NASA Astrophysics Data System was used heavily while this paper was
written. This research has made use of the NASA/IPAC Extragalactic
Database (NED) which is operated by the Jet Propulsion Laboratory,
California Institute of Technology, under contract with the National
Aeronautics and Space Administration. 
PG was partially supported during this project by NASA through grant
HST-GO-11691 from the Space Telescope Science Institute, which is operated by
the Association of Universities for Research in Astronomy, Inc., under
NASA contract NAS5--26555. 

Facilities: \facility{HST (ACS)}, \facility{Keck:I (LRIS)}, \facility{ESO:3.6m
  (EFOSC2)}

\appendix

\section{Building synthetic integrated colors of metal-rich globular
  clusters containing two stellar generations} \label{s:appendixA}

In this appendix, we determine colors of various stellar types
from synthetic spectra with chemical compositions typical of first-- and
second-generation stars found in massive metal-rich GCs. 
We assume [Z/H] = 0.0 and explore two choices for the abundances of He, C,
N, O, and Na. 
To simulate first-generation (FG) stars, we choose the primordial He
abundance \mbox{($Y$ = 0.235 + 1.5 $Z$)}, we follow \citet{cann+98} by
choosing [C/Fe] = 0.06 and [N/Fe] = 0.20, and we adopt [O/Fe] = 0.40 and
[Na/Fe] = 0.00 from \citet{carr+09}. To simulate second-generation (SG)
stars, we choose [C/Fe] = $-$0.15, [N/Fe] = 1.05, [O/Fe] = $-$0.10, and
[Na/Fe] = 0.60 \citep[see][]{carr+09}. For the He abundance of SG stars, we
consider the case of NGC~6441, a metal-rich GC in the bulge of our Galaxy
\citep[\mbox{[Fe/H] = $-$0.59, }][]{zinwes84}. NGC~6441 (as well as
the similar cluster NGC~6388) is unusual among metal-rich GCs in our
Galaxy in that it shows a horizontal branch (HB) that extends blueward
of the RR Lyrae instability strip in the CMD whereas all other
metal-rich GCs have purely red HBs
\citep[e.g.,][]{rich+97}\footnote{Note however 
  that the impact of the blue extension of the HB seen in NGC~6388 and
  NGC~6441 to the overall optical color of their HB is
  negligible. In terms of $\Delta (V\!-\!I)$, defined by
  \citet{dott+10} as the median color difference between the HB and
  the RGB at the luminosity of the HB, the averaged HB of NGC~6388 and
  NGC~6441 is 0.003 $\pm$ 0.010 mag redder than the average HB of the 8
  other Galactic GCs with Age $>$ 12 Gyr and $-0.70 < \mbox{[Fe/H]} <
  -0.50$ in that study, i.e., Lyng{\aa} 7, NGC~104, NGC~5927,
  NGC~6304, NGC~6496, NGC~6624, NGC~6637, and NGC~6838.}. The blueward
extension of the HB of NGC~6441 is widely thought to be due to
enhanced He in a fraction of its constituent stars, and we adopt the
value $Y = 0.33$ from \citet{caldan07}. As blue HBs are very unusual
among massive metal-rich GCs in our Galaxy, this value of $Y$ should
probably be considered an upper limit for SG stars in the context of
the current paper.  

\begin{figure}[tbhp]
\centerline{
\psfig{figure=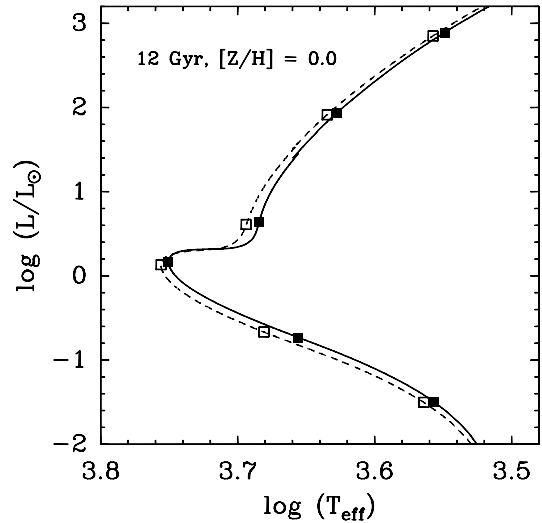,width=7.cm}
}
\caption{Theoretical isochrones from the lower MS to the upper RGB for the
  two different chemical compositions associated with two stellar generations
  within massive GCs. The solid line corresponds to 
  the first-generation population (with primordial He abundance), whereas the
  dashed line corresponds to the second-generation population (with enhanced
  He abundance $Y$ = 0.33). The filled and open squares indicate the stellar
  types for which we calculated model atmospheres and synthetic spectra (cf.\
  Table~\ref{t:atlaspars}). See text for more details. 
\label{f:isochrones}
}
\end{figure}

To produce synthetic integrated colors of stellar populations, we first
build two sets of synthetic spectra for six stellar types that span almost
the full range of luminosities $L$ and temperatures $T_{\rm eff}$ encompassed
by the isochrones (see Figure~\ref{f:isochrones}). For this purpose we select 
Dartmouth isochrones\footnote{see
  http://stellar.dartmouth.edu/~models/index.html} 
\citep{dott+07} with [Z/H] = 0.0, \aFe\ = +0.4, and the two values of $Y$
mentioned in the previous paragraph. [Z/H] values for alpha-enhanced
populations are evaluated using [Z/H] = [Fe/H] + 0.929 \aFe\ 
\citep{trag+00}. The parameters of the model atmospheres
are listed in Table~\ref{t:atlaspars}.  These temperatures, gravities, and
chemical abundances are then used to calculate model atmospheres and
synthetic spectra using the codes ATLAS12 and SYNTHE
\citep{kuru05,cast05,sbor+07}, respectively. ATLAS12 allows one to use
arbitrary chemical compositions. In doing so, we use model atmospheres from the
[Fe/H] = 0.0, \aFe\ = +0.4 model grid of \citet{caskur04}\footnote{see
  http://wwwuser.oat.ts.astro.it/castelli/grids.html} as reference and 
adjust $T_{\rm eff}$, $g$, [Fe/H], and the individual element abundances in
the ATLAS12 runs. The resulting synthetic spectra are then redshifted
according to the heliocentric radial velocity of 
the Virgo galaxy cluster \citep[1035 \kms, ][]{moul+00} and integrated over
passbands of Johnson/Cousins \citep[$U_J$, $B_J$, $V_J$, $R_C$, $I_C$; ][]{bess90} and
Sloan {\it ugriz}\footnote{see
  http://www.sdss.org/dr1/instruments/imager/index.html{\#}filters} to produce 
synthetic magnitudes and colors.   

\begin{table}[btp]
\footnotesize
\caption[]{Stellar parameters of the model atmospheres.}
\label{t:atlaspars}
\begin{center}
\vspace*{-4mm}
\begin{tabular*}{7cm}{@{\extracolsep{\fill}}cccc@{}} \tableline \tableline
\multicolumn{3}{c}{~~} \\ [-2ex]  
$T_{\rm eff}$ (FG) & $T_{\rm eff}$ (SG) & log $g$ &
 $V_{\rm turb}$ \\
(1) & (2) & (3) & (4) \\ [0.5ex] \tableline
\multicolumn{3}{c}{~~} \\ [-1.5ex] 
3602 & 3665 & 4.79 & 2.0 \\
4526 & 4795 & 4.59 & 2.0 \\
5630 & 5700 & 4.20 & 2.0 \\
4835 & 4941 & 3.48 & 2.0 \\
4237 & 4311 & 1.96 & 2.0 \\
3537 & 3608 & 0.70 & 2.0 \\ [0.5ex] \tableline 
\multicolumn{3}{c}{~~} \\ [-1.2ex]
\end{tabular*}
\end{center}
\tablecomments{Column (1): Effective temperature for first-generation
stars in K. Column (2); Effective temperature for second-generation stars in
K. Column (3): logarithm of surface gravity in cm s$^{-1}$. 
Column (4): Turbulent velocity in km s$^{-1}$.} 
\end{table}

\begin{figure}[tbp]
\centerline{
\psfig{figure=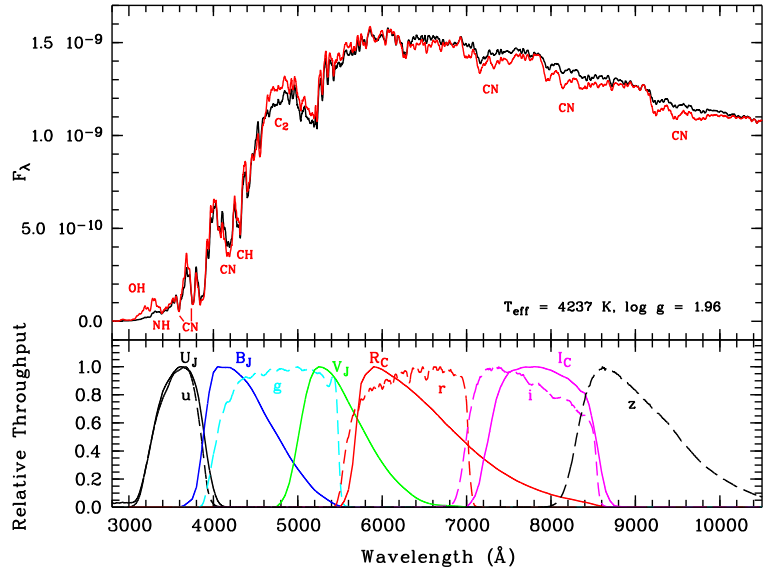,width=10cm,angle=-90.}
}
\caption{{\it Top panel:\/} Synthetic spectra for RGB models with
  log $g$ = 1.96 for the FG (black) and SG (red) mixtures. The spectra
  were convolved with a Gaussian with $\sigma$ = 12 \AA\ to improve
  readability. Relevant molecular bands are labelled. {\it Bottom
    panel:\/} Transmission curves for Johnson/Cousins $U_J$, $B_J$,
  $V_J$, $R_C$, and $I_C$ filters (solid curves, left to right) and
  SDSS {\it ugriz\/} filters (dashed curves). See text for details. 
\label{f:specplot1}
}
\end{figure}

Three examples of the synthetic spectra for FG and SG stars are shown in
Figs.~\ref{f:specplot1}\,--\,\ref{f:specplot3}. Labels in each plot
indicate prominent absorption features that change significantly in
strength between the two chemical compositions. These spectra are
described briefly below. 

Figure~\ref{f:specplot1} compares RGB spectra of the FG and SG
mixtures. The higher [N/Fe] of the SG mixture causes stronger
molecular bands of CN. This is especially clear at $\lambda \ga 7000$
\AA. The main reason why this is not as obvious for the CN bands near 3590,
3883, and 4150 \AA\ and the NH band near 3360 \AA\ is that the enhanced 
absorption in the blue region for the SG mixture is partially 
compensated by its enhanced He abundance which elevates the stellar
continuum in the blue. The increased opacity at $\lambda \la 4500$
\AA\ for the SG mixture also causes a somewhat elevated stellar
continuum at longer wavelengths, which likely causes the slightly
higher flux of the SG mixture in the 4500\,--\,6200 \AA\
region. Finally, the lower O abundance of the SG mixture causes a
weaker OH absorption band between 3050 and 3200 \AA\ which just falls
within the $U_J$ and $u$ bands for the radial velocities of the
galaxies in our sample.  

\begin{figure}[tbp]
\centerline{
\psfig{figure=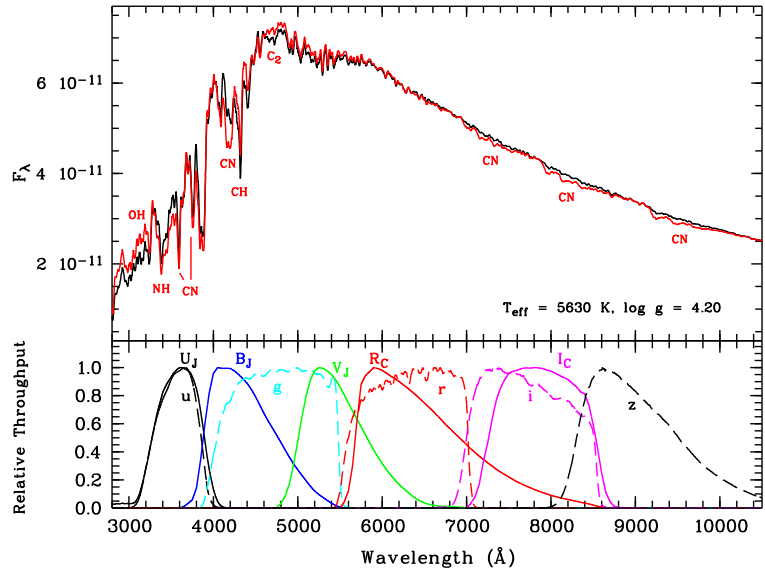,width=10cm,angle=-90.}
}
\caption{Like Figure \ref{f:specplot1}, but now for TO models with log
  $g$ = 4.20. 
\label{f:specplot2}
}
\end{figure}

\begin{figure}[tbp]
\centerline{
\psfig{figure=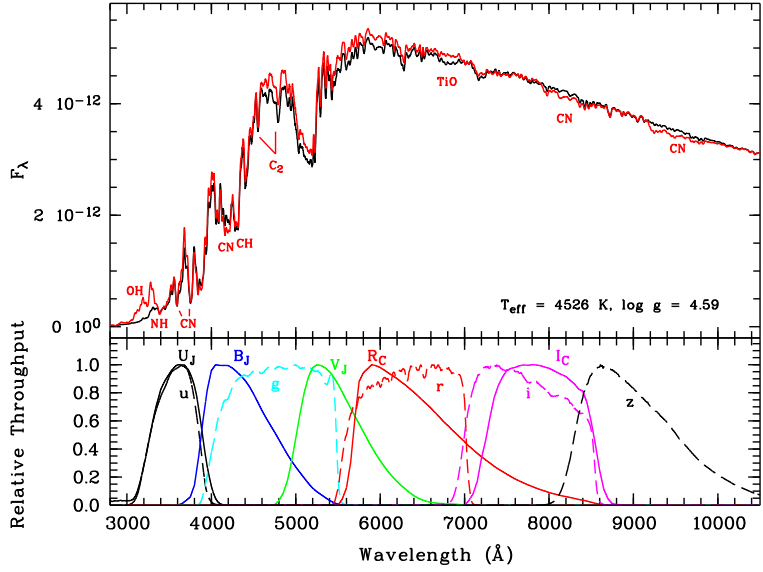,width=10cm,angle=-90.}
}
\caption{Like Figure \ref{f:specplot1}, but now for MS models with log
  $g$ = 4.59. 
\label{f:specplot3}
}
\end{figure}

In the spectra of TO stars (Figure~\ref{f:specplot2}), the higher
temperature causes generally weaker molecular bands, rendering
generally small differences between the FG and SG spectra. Only the NH
and CN bands remain slightly stronger in the SG mixture. 

The differences between the FG and SG spectra of cool MS stars
(Figure~\ref{f:specplot3}) are generally similar to those seen for the
RGB spectra, except that the MS stars show weaker red CN bands and
a slightly stronger OH band in the 3000\,--\,3200 \AA\ region.

\begin{figure}[tbh]
\centerline{
\psfig{figure=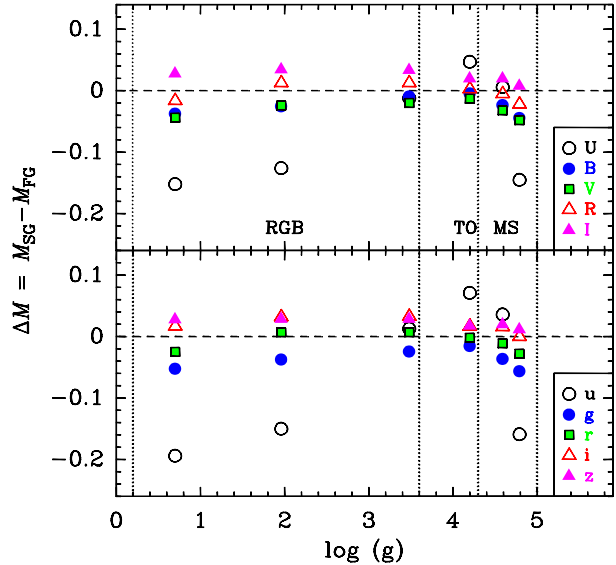,width=8.cm}
}
\caption{Differences between stellar absolute magnitudes of the 
  FG and SG stars considered in this paper as a function of log\,$g$. {\it 
    Upper panel:\/} Johnson/Cousins {\it UBVRI\/} filters (see legend for
  symbols). {\it Lower panel:\/} Sloan {\it ugriz\/} filters. Dashed
  vertical lines delineate the RGB, TO, and MS regions on the CMD (see labels
  near the bottom of the upper panel). See Appendix~\ref{s:appendixA} for
  more details.   
\label{f:deltaBCs}
}
\end{figure}

Differences in stellar absolute magnitude $\Delta M \equiv M_{\rm SG} -
M_{\rm FG}$ are then evaluated as a function of log $g$
for all filter passbands considered here. The results are illustrated in
Figure~\ref{f:deltaBCs} which shows $\Delta M$ as a function of log $g$. 
The largest differences show up in the $U_J$ and $u$ passbands, due to
their sensitivity to relevant parameters: {\it (i)\/} $T_{\rm eff}$
and hence $Y$; {\it (ii)\/} N abundance through the NH band near 3360 
\AA\ as well as the CN bands near 3590 and 3883 \AA; and {\it (iii)\/} O
abundance through the OH band at the short-wavelength edge of the
$U_J$ and $u$ filters. 
The weaker OH absorption for the SG population is the main reason
why $\Delta M$ for the $U_J$ and $u$ passbands is negative at the top
of the RGB (i.e., lowest log\,$g$). $\Delta M$ then increases
significantly with increasing log $g$ to become positive at the TO
(due mainly to stronger NH and CN absorption in the SG population),
and then decreases again to the bottom of the MS (i.e., highest log\,$g$).   
The other passbands (longward of $\sim 4000$ \AA) show a similar
dependence of $\Delta M$ on log\,$g$, although the absolute $\Delta M$ values
at $\lambda \ga 4000$ \AA\ are much smaller than for $U_J$ and $u$ at a given
log\,$g$. 

The dependence of $\Delta M$ between FG and SG populations as a
function of wavelength found here are qualitatively similar to those of
\citet{sbor+11} who did a similar study at low metallicity
([Fe/H] = $-$1.62). However, we find that $\Delta M$ values are
different for metal-rich populations in a quantitative sense,
sometimes by a significant amount. 

Isochrone tables for the SG population are then calculated by adding $\Delta
M$ to the absolute magnitudes for each filter passband in the (FG) isochrone
tables (i.e., $M_{\rm SG} = M_{\rm FG} + \Delta M$), using linear interpolation
in log $g$. After rebinning the resulting isochrone tables to a uniform bin
size in  stellar mass using linear interpolation, weighted integrated
luminosities of the population were derived for the relevant passbands by
weighting individual stellar luminosities in the isochrones using a
``standard'' \citet{krou01} IMF.  Resulting integrated-light magnitude offsets
(in the sense SG $-$ FG) are listed for each filter passband in 
Table~\ref{t:intmags}. 
Interestingly, the large differences found in the $U_J$ and $u$ passbands
between FG and SG populations at a {\it given\/} stellar type largely cancel
out in {\it integrated\/} light (i.e., when integrated over the stellar mass
function). Integrated-light magnitude offsets are found to stay within 0.03
mag in an absolute sense for any of the filter passbands considered
here. Finally, and in the context of the subject of this paper, we note that the
resulting offsets between SG and FG populations in the \BI\ and \gz\
colors are $-$0.041 and $-$0.054 mag (in the sense SG\,$-$\,FG), respectively. 
Since recent multi-object spectroscopy studies of massive Galactic GCs
typically show an approximate 50\,--\,50\% split between FG and SG
stars \citep[e.g.,][]{grat+12}, we conclude that the predicted overall effect
of the presence of multiple stellar generations on the integrated \BI\ and
\gz\ colors of massive metal-rich GCs is offsets of $-$0.021 and $-$0.027
mag, respectively.  

\begin{table}[thbp]
\footnotesize
\caption[]{Integrated-light magnitude offsets between the SG and FG populations.}
\label{t:intmags}
\begin{center}
\vspace*{-4mm}
\begin{tabular*}{8.cm}{@{\extracolsep{\fill}}ccccc@{}} \tableline \tableline
\multicolumn{3}{c}{~~} \\ [-2ex]  
$\Delta U_J$ & $\Delta B_J$ & $\Delta V_J$ & $\Delta R_C$ & $\Delta I_C$ \\ [0.5ex] \tableline
\multicolumn{3}{c}{~~} \\ [-1.5ex] 
$+$0.002 & $-$0.018 & $-$0.027 & $-$0.002 & $+$0.023 \\ [0.5ex] \tableline \tableline
\multicolumn{3}{c}{~~} \\ [-2ex]  
$\Delta u$ & $\Delta g$ & $\Delta r$ & $\Delta i$ & $\Delta z$ \\ [0.5ex] \tableline
\multicolumn{3}{c}{~~} \\ [-2ex]
 $+$0.022 & $-$0.031 & $-$0.007 & $+$0.018 & $+$0.023 \\ [0.5ex] \tableline 
\multicolumn{3}{c}{~~} \\ [-1.8ex]
\end{tabular*}
\tablecomments{Magnitude offsets are in the sense SG $-$ FG.}
\end{center}
\end{table}











\end{document}